\documentclass[aps,prl,twocolumn,superscriptaddress]{revtex4}

\usepackage{graphicx}

\begin{document}

%\preprint{}

\title{Absence of a large superconductivity-induced  gap in magnetic fluctuations of Sr$_2$RuO$_4$}


\author{S. Kunkem\"oller}
\affiliation{$I\hspace{-.1em}I$. Physikalisches Institut,
Universit\"at zu K\"oln, Z\"ulpicher Str. 77, D-50937 K\"oln,
Germany}

\author{P. Steffens}
\affiliation{Institut Laue Langevin,71 avenue des Martyrs,
38000 Grenoble, France}

\author{P. Link}
\affiliation{Heinz Maier-Leibnitz Zentrum, Technische
Universit\"at M\"unchen, Lichtenbergstrasse 1, 85748 Garching,
Germany}

\author{Y. Sidis}
\affiliation{Laboratoire L\'eon Brillouin, C.E.A./C.N.R.S.,
F-91191 Gif-sur-Yvette CEDEX, France}

\author{Z. Q. Mao}
\affiliation{Department of Physics, Graduate School of Science,
Kyoto University, Kyoto 606-8502, Japan}
\affiliation{Department of Physics, Tulane University, New Orleans, LA 70118, USA}

\author{Y. Maeno}
\affiliation{Department of Physics, Graduate School of Science,
Kyoto University, Kyoto 606-8502, Japan}

\author{M. Braden}\email[e-mail: ]{braden@ph2.uni-koeln.de}
\affiliation{$I\hspace{-.1em}I$. Physikalisches Institut,
Universit\"at zu K\"oln, Z\"ulpicher Str. 77, D-50937 K\"oln,
Germany}

%\thanks{}


%\date{\today}

\begin{abstract}

Inelastic neutron scattering experiments on Sr$_2$RuO$_4$
determine the spectral weight of the nesting induced magnetic
fluctuations across the superconducting transition. There is no
observable change at the superconducting transition down to an
energy of $\sim$0.35~meV, which is well below the 2$\Delta$ values
reported in several tunneling experiments. At this and higher
energies magnetic fluctuations clearly persist in the
superconducting state. Only at energies below $\sim$0.3~meV evidence for partial
suppression of spectral weight in the superconducting state can be
observed. This strongly suggests that the one-dimensional bands
with the associated nesting fluctuations do not form the active,
highly gapped bands in the superconducting pairing in
Sr$_2$RuO$_4$.

\end{abstract}

\pacs{7*******}

%\keywords{}

\maketitle

Sr$_2$RuO$_4$ is one of the best studied unconventional superconductors  \cite{1,2,3,4,5} but its pairing symmetry and mechanism still remain a subject of very active debate. There is newly added evidence in favor of the most advocated symmetry of the superconducting order, namely the spin-triplet chiral p-wave symmetry, such as the increase in the Knight shift expected in the equal-spin-pairing (ESP) triplet state \cite{6}, observation of the surface density of states consistent with the chiral edge state \cite{7}, and the magnetization steps corresponding to the half-quantum fluxoids \cite{8}. On the other hand, there are results challenging the p-wave pairing scenario, such as the strong limiting of the in-plane upper critical fields \cite{9}, the first-order superconducting transition \cite{10,11}, and the absence of the chiral edge current \cite{12}. At present, there seems no symmetry model which can explain all the experimental facts available. If the most
advocated symmetry of the superconducting order is correct, Sr$_2$RuO$_4$ would be a topological superconductor proposed as a promising candidate for quantum computing \cite{quancom1,quancom2}.

Another prominent feature of Sr$_2$RuO$_4$ is that its normal state is quantitatively well characterized as a quasi-two-dimensional (Q2D) Fermi liquid \cite{2,3}. The Fermi surface consists of three cylindrical sheets \cite{2}: two originate from the $d_{xz}$ and $d_{yz}$ orbitals, called the $\alpha$ and $\beta$ bands, and retain a quasi-one-dimensional (Q1D) character as well; the other one from the $d_{xy}$, called the $\gamma$ band, shows a Q2D character. All three bands disperse weakly along the interlayer $c$ direction \cite{13}. In such a multiband system with distinct orbital symmetries, superconductivity may be strongly orbital dependent \cite{14}. The strong nesting between the Q1D bands results in strongly enhanced spin-density wave (SDW) fluctuations \cite{15,16,17,18,19,20} and even minor chemical substitution leads to static ordering of this SDW instability with the moment along the c direction. Only 2.5\% of Ti induce this SDW phase \cite{21,22}, and recent muSR experiments and neutron scattering studies show that the same magnetic order occurs upon replacing Sr with isovalent Ca \cite{23,24}. Such spin fluctuations originating from the nesting of the Q1D Fermi surface sheets cannot easily lead to the most likely chiral superconducting state \cite{2}.
The equal-spin p-wave pairing scenario is based on quasi-ferromagnetic correlations associated with the $\gamma$ band, and amongst the various p-wave possibilities a chiral (and topological) state, $k_x + ik_y$, was proposed to explain various experiments \cite{2,3}. Evidence for strong quasi-ferromagnetic fluctuations can be found in
susceptibility \cite{2,3} and NMR measurements \cite{nmr}, but a thorough study of such fluctuations still lacks.
Thus, one important step toward resolving the apparent controversy is to identify which of the bands are mainly responsible for the superconductivity.

%=========================================================
\begin{figure}
\includegraphics[width=0.85\columnwidth]{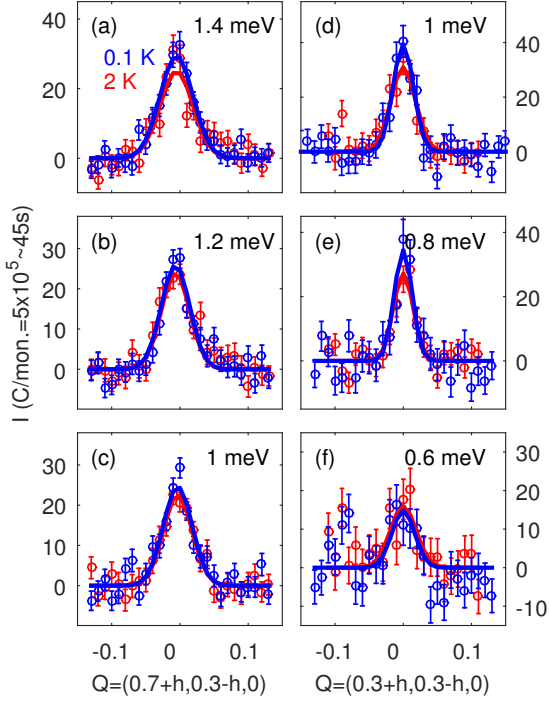}
\caption{\label{tdep} Constant-energy scans obtained on THALES
with $k_f=1.57$~\AA$^{-1}$ using the PG-PG configuration. Intensity profiles were
fitted by the sum of a Gaussian peak and a curved background, which was assumed
identical at both temperatures and subtracted from the data.
}
\end{figure}
%=========================================================

Many attempts were made to reconcile the discrepancy between the
pairing symmetry and the apparently dominant magnetic fluctuations
\cite{2,3,5}. Treating the on-site Coulomb repulsion within perturbation theory corroborates the scenario of $p$-wave pairing mainly arising in
the Q2D band \cite{nomura}. This scenario is challenged by Raghu et al., who apply renormalization group
techniques and discuss orbital and charge fluctuations in the
Q1D bands as the main ingredient \cite{raghu}. These calculations
were extended by Scaffidi et al. \cite{scaffidi14} to include inter-band and spin-orbit
coupling yielding similar sized gaps on all bands without tuning of parameters.
In contrast the recent analysis by Huo,
Rice and Zhang argues in favor of superconductivity arising in the
Q2D bands with the nesting fluctuations perturbing the superconductivity
\cite{huo}.  Experimentally, the observation
of a strong enhancement of the superconducting $T_c$ (by a factor
~2!) under both tensile and compressive strain \cite{hicks} may suggest
a dominant influence of the van Hove singularity in the Q2D bands
associated with the ferromagnetic instability. The question which
bands drive superconductivity in Sr$_2$RuO$_4$ remains as open and
fascinating as ever \cite{kallin}.

Inelastic neutron
scattering (INS) can yield valuable information concerning the
role of the different bands in the pairing \cite{huo}. If superconductivity
directly arises from the Q1D bands as active bands, which thus exhibit a large gap,
there must be a clear impact on the associated incommensurate
magnetic excitations. Several calculations explicitly predict the occurrence
of a resonance mode in at least one of the spin excitation
channels for $p$ wave superconducting symmetry \cite{huo,kee,morr,yakiyama}. On the other hand, if
superconductivity is mainly driven by the Q2D band associated with
ferromagnetic fluctuations, a lower gap in the Q1D bands and only a
small impact on the magnetic fluctuations is expected \cite{14,huo}. Here we
report INS experiments across the superconducting transition in
Sr$_2$RuO$_4$, which clearly show that nesting-induced magnetic
fluctuations only sense a very small gap suggesting that the Q1D bands
are not the active ones in the superconducting pairing.

%=========================================================
\begin{figure}
\includegraphics[width=0.85\columnwidth]{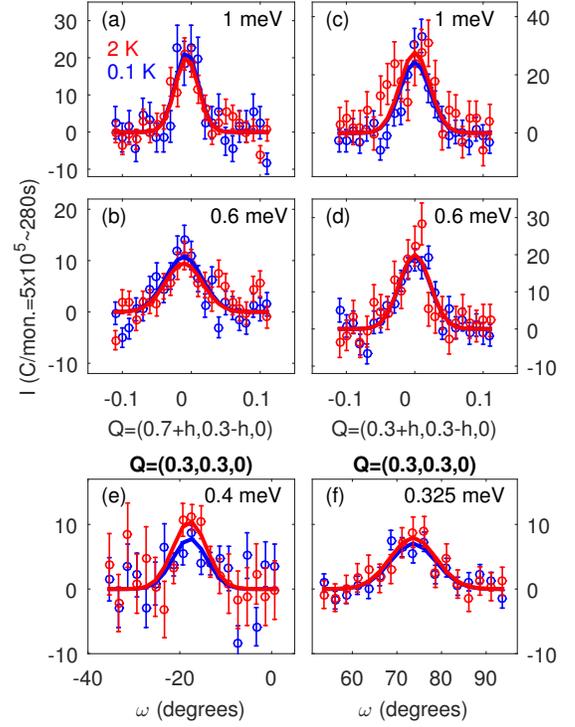}
\caption{\label{fig2} (a-d) Constant-energy scans obtained on THALES
with $k_f=1.57$~\AA$^{-1}$ using the Si-PG configuration.
A flat background was subtracted from the data. (e-f) Inelastic rocking scans 
using the Si-PG configuration. The
sample was rotated through the $(0.3,0.3,0)$ $E=0.4$ and 0.325~meV
positions at the center of the scans yielding a flat background that was subtracted
from the data.
}
\end{figure}
%=========================================================

The difficulty of INS experiments on the magnetic response in the
superconducting state of Sr$_2$RuO$_4$ consists in the weakness of
the signal combined with the high resolution needed. The INS
intensity is given by the imaginary part of the generalized
susceptibility, $\chi''({\bf Q},E)$, multiplied by the Bose factor \cite{17}:
\begin{equation}
\frac{d^2\sigma}{d\Omega dE}=\frac{k_f{r_0}^2F^2(Q)}{k_i
\pi(g\mu_B)^2}\frac{2\cdot\chi''({\bf Q},E)}{1-exp(-E/k_bT)},
\end{equation}
where we ignore the spin anisotropy of the magnetic susceptibility \cite{19}
($k_i$ and $k_f$ denote incoming and final neutron momentum,
$F(Q)$ the magnetic form factor of Ru at the scattering vector and
${r_0}^2=0.29\cdot10^{-28}$~m$^2$). The nesting-induced
magnetic excitations at ${\bf q}_{inc}$ follow a single relaxor behavior
\cite{16,17,19,20}:
\begin{equation}
 \chi''({\bf q}_{inc},E)=\chi'({\bf q}_{inc},0)\frac{\Gamma E}{\Gamma ^2 +
E^2},\end{equation} which is maximum at the characteristic energy
$\Gamma$ and almost linear for much lower energies. INS
experiments in the normal state indicate strong magnetic
scattering at the nesting vector, ${\bf q}_{inc}$, with the characteristic
energy decreasing towards low temperatures. But this softening
stops at $\Gamma$$\sim$6~meV, which is well
above the values of the superconducting gap \cite{16,17,19}.
Therefore, the INS signal in the range, where one may expect an
impact of the superconducting gap, is very small. In addition, the
experiment requires a high energy resolution in order to study
this region close to the strong elastic response, which
considerably reduces the INS intensity. Due to these difficulties the
previous INS experiments on Sr$_2$RuO$_4$ in the superconducting
phase yielded reliable statistics on the nesting fluctuations only
for energy transfer above $\sim$1~meV \cite{17}.

INS experiments were carried out on the PANDA triple-axis
spectrometer at the Forschungsreaktor Munich II and at the
recently upgraded THALES instrument at the Institut Laue Langevin.
In all experiments we used an assembly of 12 Sr$_2$RuO$_4$
crystals with a total volume of 2.2~cm$^3$. The crystals were grown
at Kyoto University using a floating-zone image furnace
and similar crystals were studied in many different experiments
\cite{2,3}. We choose the [100]/[010] scattering geometry, because
this yields the best INS signal due to the integration along the
vertical direction along $c$ where little modulation of magnetic
response is expected. For all experiments the crystal assembly
was cooled with a dilution refrigerator attaining minimum temperature
of the order of $\sim$50~mK. There is some impact on the neutron
absorption on the sample temperature of the order of 10~mK,
which, however, is negligible compared to the transition
temperature. On PANDA we mostly used a final momentum of
$k_f=1.2$~\AA $^{-1}$ to obtain sufficient resolution and pyrolitic graphite (PG) (002)
as monochromator and analyzer. In order to
decrease the background a BeO filter was put in front of the
analyzer and a Be filter between monochromator and sample. On THALES a much better intensity
to background ratio was achieved, but some
residual background at low energies remained when using PG
(002) monochromator and analyzer crystals  (PG-PG configuration) even for rather small
values of the final momentum. In order to further suppress this low-energy background
we included a radial collimator and a Be filter in front of the
analyzer and we used a Si (111) monochromator (SI-PG configuration). We applied vertical and horizontal
focusing at both the monochromator and analyzer. In
addition, a velocity selector in front of the monochromator was
inserted to suppress higher order contaminations. Most scans on
THALES were performed with a fixed final momentum of $k_f=1.57$~\AA
$^{-1}$ where the Be filter effectively cuts all neutrons with
only slightly larger final energy. Some scans
were performed by scattering at the sample and at the analyzer in
the same sense (U configuration), which reduces the background as
the detector is positioned farther away from the direct beam, but
slightly worsens the resolution.

In spite of serious efforts the measurements on PANDA considerably
suffered from the background scattering. Scans at the scattering
vectors of $(0.3,0.3,0)$ and $(0.7,0.7,0.4)$ did not yield any
indication for a superconductivity induced change at $T_c$ above
E$\sim$0.6~meV but the achieved statistics at lower energy remained
insufficient to characterize the weak magnetic signal. In the
following we will therefore focus on the results obtained on
THALES which exhibit significantly better statistics.

Figures 1 and 2 show constant energy scans 
for temperatures above and below the superconducting transition.
The data in Fig. 1 were taken with the PG-PG configuration on THALES
(energy resolution at the elastic line $\Delta E_0=0.20$~meV full width at half maximum)
and those in Fig. 2 with the SI-PG configuration, which yields a lower background at small energies and improves the resolution ($\Delta E_0=0.12$~meV) but
considerably reduces the signal. With
the dilution refrigerator cryostat used in these experiments, it is
not possible to obtain sufficient temperature stability
in the range 1.2 to 1.6~K, therefore we
could not follow the signals close to $T_c$. The data shown in
Fig. 1 and Fig. 2 unambiguously show that the nesting related
fluctuations in the energy range 0.6 to 1~meV can be easily studied
by our INS experiment and that this signal is not affected by the
superconducting transition concerning neither the intensity nor
the width. We have studied the nesting signal at the two
scattering vectors ${\bf Q}=(0.3,0.3,0)$ and $(0.7,0.3,0)$ which are
not equivalent due to the centering of the body centered lattice
in Sr$_2$RuO$_4$ and due to the lower form factor at the latter
reducing the magnetic signal. Because of the quasi-twodimensional
nature of the magnetic correlations in Sr$_2$RuO$_4$, however, one
does not expect an essential difference, and the signal at both
scattering vectors is comparable and in particular there is no
change at the superconducting transition, $T_c=1.4$~K for energies above 0.6~meV at both ${\bf Q}$ values.

Experiments at lower energy transfer are more difficult as
described above. Since the background depends on the length of the
scattering vector (i.e. the scattering angle), it is not constant in
a straight transversal constant-energy scan like those shown in
Fig. 1 and 2(a-d) but may peak at the scan center.
Therefore, we performed inelastic rocking scans
by turning the sample with fixed $\vert \textbf{Q} \vert$, see Fig. 2(e-f).
These scans posses a flat background and clearly confirm that
magnetic scattering persists in the superconducting state essentially unchanged down
to energies of the order of 0.325~meV. Note that the Bose factor
explains a small intensity reduction between 2 and 0.1~K of
1.18 and 1.11 at $E=0.325$ and 0.4~meV respectively, so that the data do not
yield any significant reduction of spectral weight even at 0.325~meV.

%=========================================================
\begin{figure}
\includegraphics[width=0.74\columnwidth]{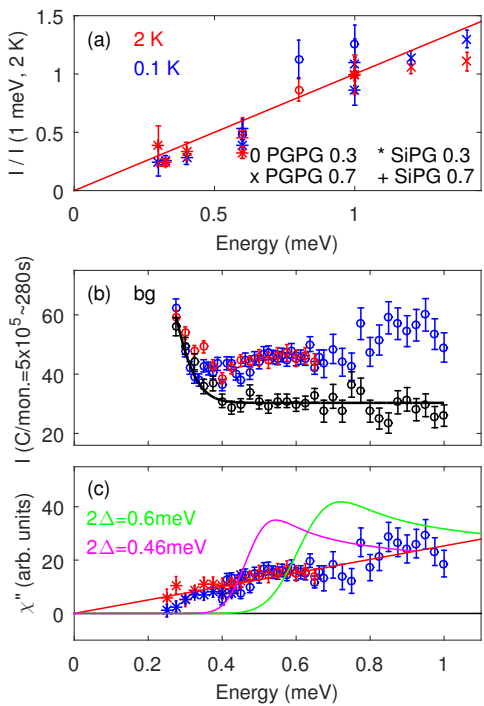}
\caption{\label{fig4} (a) Fitted Gaussian peak heights
obtained from the constant-$E$ scans taken
with the two configurations at the two scattering vectors. In
order to allow for comparison, the data were normalized to the values
at 1~meV and 2~K and  a correction for the Bose factor was applied.
 (b) Constant-${\bf Q}$ scans obtained with
$k_f=1.57$~\AA$^{-1}$ using the Si-PG configuration. Blue and red
symbols denote the data taken at ${\bf Q}=(0.3,0.3,0)$  above and below
the superconducting transition, respectively, and black symbols
denote background intensity observed at a ${\bf Q}$ vector of the same length but
rotated by 16$^\circ$ with respect to the ${\bf Q}$ position of the
nesting response. (c) Magnetic signal at ${\bf Q}=(0.3,0.3,0)$ obtained by subtracting the
background signal and by correcting for the Bose factor.  Straight lines in (a) and (c) denote the linear relation
$ \chi'' \propto E $ expected for the single relaxor at low energy, see equation (2).
The magenta and green lines in (c) correspond to the calculated magnetic response in the case of
active Q1D bands \cite{huo}, which results in a resonance excitation at the energy of twice the
superconducting gap taken at the weak coupling BCS value 0.46~meV and at the value
observed in tunneling experiments (0.6~meV, see text). The theoretical result was 
folded with the experimental resolution.
Data in (b) and part of the data in (c) (circles) were taken in U-configuration yielding lower background at energies above 0.4~meV and slightly reduced energy resolution, $\Delta E_0=0.16$~meV, while the low-energy part of (c) was recorded in z-configuration (stars) in a dedicated experiment. In this z-configuration the background remains flat and a better statistics could be reached resulting in much smaller error bars.}
\end{figure}
%=========================================================

Figure 3 resumes the energy dependence of the magnetic nesting signal.
Figure 3(a) shows the fitted peak heights of the
constant-$E$ scans taken in different configurations at the two scattering
vectors. In order to compare data taken at different ${\bf Q}$ positions, in
different configurations (scattering sense at the analyser) and in different runs, intensities are normalized to
the values at 1~meV and 2~K. The peak heights at larger energies
remain unchanged upon entering the superconducting state
while evidence for partial suppression of spectral weight is observed below $\sim$0.3~meV. Figure 3(b)
shows constant ${\bf Q}$-scans taken at the nesting
scattering vector $(0.3,0.3,0)$ above and below the superconducting
transition as well as a background scan taken at a scattering vector of the
same length but rotated 16 degrees away from the nesting
position. Subtracting this background signal from that obtained
at the nesting ${\bf Q}$-position we can deduce the magnetic
signal at both temperatures, see Fig. 3(c). This analysis shows that the nesting
scattering remains essentially unchanged for energies above $\sim$0.325~meV. The constant Q-scans data yield week evidence for partial suppression of spectral weight due to the
opening of the superconducting gap only at very low energies, see Fig. 3(c), but additional studies are desirable.

The magnetic response of an itinerant system corresponds to a
particle-hole excitation, which in a superconductor must cross twice the superconducting gap, 2$\Delta$. There have
been several reports on the superconducting gap in Sr$_2$RuO$_4$
\cite{15,laube,upward,suderow,7}: The first tunneling
experiments were interpreted as evidence for very large gap and
$\frac{2\Delta}{k_BT_c}$ values \cite{laube,upward} while more
recent studies conclusively suggest smaller values: Suderov
et al. $2\Delta=0.56$~meV \cite{suderow}, Kashiwara et al.
$2\Delta=0.93$~meV \cite{7} and Firmo et al.
$2\Delta=0.7$~meV slightly above the weak coupling BCS value $2\Delta=0.46$~meV. None of the tunneling studies can safely identify the
band carrying the largest gap, leaving the discussion about active
and passive bands open. On the theoretical side, different
studies arrive at nearly the same conclusion that opening the $p$-wave gap
in the Q1D sheets results in a full suppression of spectral weight
below $2\Delta_{1d}$ and even a resonance enhancement at or close
to this value \cite{12,kee,morr,yakiyama}. In Fig. 3(c) we include
the calculation for a superconducting gap opening in the Q1D bands
of 0.46 and 0.6~meV \cite{huo} folded with the experimental resolution. 
Our results clearly contradict such picture.
A resonance enhancement of
the magnetic response in the superconducting state has been
reported in several unconventional superconductors \cite{scalapino}.
In particular in superconductors, in which the pairing appears
mediated by well defined magnetic fluctuations such as the cuprates or
the FeAs-based compounds, strong resonance modes are found
\cite{scalapino}. Such a behavior can be excluded for the nesting
scattering in Sr$_2$RuO$_4$ which exhibits no significant suppression
of magnetic weight at energies well below the
maximum $2\Delta$ reported in the tunneling experiments or the weak coupling BCS value.
It seems therefore very unlikely that the Q1D bands are the active ones
for the superconducting pairing in Sr$_2$RuO$_4$.
Instead the ferromagnetic fluctuations arising from the large density of states in the
Q2D bands can imply superconductivity primordially  in the Q2D bands. This scenario is
supported by the field-orientation dependence of the specific heat \cite{deguchi1,deguchi2} and NMR data \cite{2,3,5,kallin}, and direct evidence for ferromagnetic
fluctuations can be obtained from magnetization \cite{2,3} and polarized INS studies \cite{bradenpol}.

Nodes of the gap function may lead to persisting magnetic
scattering in the superconducting state for energies below the
maximum values of $2\Delta$. But in the scenario of Q1D bands being
the active ones for the superconducting pairing mediated by
nesting induced fluctuations, some effect of the gap opening must
be observed. The fact that there is no change in the magnetic
scattering (below 20\% for $E$$>0.325$~meV) well below the observed
maximum values of $2\Delta$ \cite{7,suderow,laube,upward,
15} renders such a scenario very unlikely.

In conclusion we have studied the low-energy magnetic fluctuations
associated with the nesting of Q1D bands in Sr$_2$RuO$_4$. The fact
that we do not observe a significant change in this signal when
passing the superconducting transition disagrees with a scenario
of nesting related fluctuations driving superconductivity
primordially in the Q1D bands.

%\begin{acknowledgments}
This work was supported by the Deutsche Forschungsgemeinschaft
through  CRC 1238 Project No.  B04 and by the JSPS KAKENHI No. JP15H05852.


\begin{thebibliography}{9}


\bibitem{1} Y. Maeno, H. Hashimoto, K. Yoshida, S. Nishizaki, T. Fujita, J. G. Bednorz and F. Lichtenberg, Nature (London) {\bf 372}, 532 (1994).

\bibitem{2} A. P. Mackenzie and Y. Maeno, Rev. Mod. Phys. {\bf 75}, 657 (2003).

\bibitem{3} Y. Maeno, S. Kittaka, T. Nomura, S. Yonezawa, and K. Ishida, J.
Phys. Soc. Jpn. {\bf 81}, 011009 (2012).

\bibitem{4} Y. Liu and Z.-Q. Mao, Physica (Amsterdam) {\bf 514C}, 339 (2015).

\bibitem{5} C. Kallin and A. Berlinsky, Rep. Prog. Phys. {\bf 79}, 054502 (2016).

\bibitem{6} K. Ishida, M. Manago, T. Yamanaka, H. Fukazawa, Z. Q. Mao, Y. Maeno, and K. Miyake, Phys. Rev. B {\bf 92}, 100502(R) (2015).

\bibitem{7} S. Kashiwaya, H. Kashiwaya, H. Kambara, T. Furuta, H. Yaguchi, Y. Tanaka, and Y.
Maeno, Phys. Rev. Lett. {\bf 107}, 077003 (2011).

\bibitem{8} J. Jang, D. G. Ferguson, V. Vakaryuk, R. Budakin, S. B. Chung, P. M. Golbart, and Y. Maeno, Science {\bf 331}, 186 (2011).

\bibitem{9} C. Rastovski, C. D. Dewhurst, W. J. Gannon, D. C. Peets, H. Takatsu, Y. Maeno, M. Ichioka, K. Machida, and M. R. Eskildsen, Phys. Rev. Lett. {\bf 111}, 087003 (2013).

\bibitem{10} S. Yonezawa, T. Kajikawa, and Y. Maeno, Phys. Rev. Lett. {\bf 110}, 077003 (2013).

\bibitem{11}  S. Kittaka, A. Kasahara, T. Sakakibara, D. Shibata, S. Yonezawa, Y. Maeno, K. Tenya, and K. Machida, Phys. Rev. B {\bf 90}, 220502(R) (2014).

\bibitem{12}  C. W. Hicks, J. R. Kirtley, T. M. Lippman, N. C. Koshnick, M. E. Huber, Y. Maeno, W. M. Yuhasz, M. B. Maple, and K. A. Moler, Phys. Rev. B {\bf  81}, 214501 (2010).

\bibitem{quancom1} D. A. Ivanov, Phys. Rev. Lett. {\bf 86}, 268 (2001).

\bibitem{quancom2} C. Nayak, S. H. Simon, A. Stern, M. Freedman, and S. D. Sarma, Rev. Mod. Phys. {\bf 80}, 1083 (2008).

\bibitem{13} C. N. Veenstra, Z.-H. Zhu, M. Raichle, B. M. Ludbrook, A. Nicolaou, B. Slomski, G. Landolt, S. Kittaka, Y. Maeno, J. H. Dil, I. S. Elfimov, M. W. Haverkort, and A. Damascelli, Phys. Rev. Lett. {\bf 112}, 127002 (2014).

\bibitem{14} D.F. Agterberg, T.M. Rice, and M. Sigrist, Phys. Rev. Lett. {\bf 78}, 3374 (1997).

\bibitem{15} I. A. Firmo, S. Lederer, C. Lupien, A. P. Mackenzie, J. C. Davis, and S. A. Kivelson, Phys. Rev. B {\bf 88},
134521 (2013).

\bibitem{16}  Y. Sidis, M. Braden, P. Bourges, B. Hennion, S. NishiZaki,
Y. Maeno, and Y. Mori, Phys. Rev. Lett. {\bf 83}, 3320 (1999).

\bibitem{17} M. Braden, Y. Sidis, P. Bourges, P. Pfeuty, J. Kulda, Z. Mao,and
Y. Maeno, Phys. Rev. B {\bf 66}, 064522 (2002).

\bibitem{18} F. Servant, B. Fak, S. Raymond, J. P. Brison, P. Lejay, and J.
Flouquet, Phys. Rev. B {\bf 65} , 184511 (2002).

\bibitem{19} M. Braden, P. Steffens, Y. Sidis, J. Kulda, P. Bourges, S. Hayden,
N. Kikugawa, and Y. Maeno, Phys. Rev. Lett. {\bf 92}, 097402
(2004).

\bibitem{20} K. Iida, M. Kofu, N. Katayama, J. Lee, R. Kajimoto, Y. Inamura, M.
Nakamura, M. Arai, Y. Yoshida, M. Fujita, K. Yamada, and S.-H. Lee
Phys. Rev. B {\bf 84}, 060402(R) (2011).

\bibitem{21} M. Minakata and Y. Maeno, Phys. Rev. B {\bf 63}, 180504(R) (2001).

\bibitem{22} M. Braden, O. Friedt, Y. Sidis, P. Bourges, M. Minakata and Y. Maeno,  Phys. Rev. Lett. {\bf 88}, 197002 (2002).


\bibitem{23} J.P. Carlo, T. Goko, I. M. Gat-Malureanu, P. L. Russo, A. T. Savici, A. A. Aczel, G. J. MacDougall, J. A. Rodriguez, T. J. Williams, G. M. Luke, C. R. Wiebe, Y. Yoshida, S. Nakatsuji, Y. Maeno, T. Taniguchi and Y. J. Uemura, Nat.  Mater. {\bf 11}, 323 (2012).

\bibitem{24} S. Kunkem\"oller, A. A. Nugroho, Y. Sidis and M. Braden, Phys. Rev. B {\bf 89}, 045119 (2014)

\bibitem{nmr} K. Ishida, H. Mukuda, Y. Minami, Y. Kitaoka, Z. Q. Mao, H. Fukazawa, and Y. Maeno,
Phys. Rev. B  {\bf 64}, 100501 (2001).


\bibitem{nomura} Y. Yanase, T. Jujo, T. Nomura, H. Ikeda,
T. Hotta, and K. Yamada, Phys. Rep. {\bf 387}, 1 (2003).


\bibitem{raghu} S. Raghu, A. Kapitulnik, and S. A. Kivelson, Phys. Rev. Lett. {\bf 105}, 136401 (2010).



\bibitem{scaffidi14} T. Scaffidi, J. C. Romers, and S. H. Simon, Phys. Rev. B {\bf 89}, 220510(R) (2014)
\bibitem{huo} J.W. Huo, T.M. Rice, and F.-C. Zhang, Phys. Rev. Lett. {\bf 110}, 167003 (2013).

%\bibitem{scaffidi15}
%T. Scaffidi and S. H. Simon, Phys. Rev. Lett. {\bf 115}, 087003
%(2015).

\bibitem{hicks} C. W. Hicks, D. O. Brodsky, E. A. Yelland, A. S. Gibbs, J. A. N. Bruin, M. E. Barber, S. D. Edkins, K. N. Nishimura, S. Yonezawa, Y. Maeno, and A. P. Mackenzie, Science {\bf 344}, 283 (2014).

\bibitem{kallin} C. Kallin, Rep. Prog. Phys. {\bf 75}
 (2012).

\bibitem{laube} F. Laube, G. Goll, H. v. L\"ohneysen, M. Fogelstr\"om, and F.
Lichtenberg,
Phys. Rev. Lett. {\bf 84}, 1595 (2000).

\bibitem{upward}M. D. Upward, L. P. Kouwenhoven, A. F. Morpurgo, N. Kikugawa, Z. Q. Mao, and Y.
Maeno, Phys. Rev. B {\bf 65}, 220512(R) (2002).

\bibitem{suderow} H. Suderow, V. Crespo, I. Guillamon, S. Vieira,
F. Servant, P. Lejay, J. P. Brison, and J. Flouquet, New J.
Phys. {\bf 11}, 093004 (2009).

%\bibitem{kashiwaya}
\bibitem{kee} Hae-Young Kee, J. Phys, Condens. Matter {\bf 12},
2279  (2000).

\bibitem{morr} D. K. Morr, P. F. Trautman, and M. J. Graf, Phys. Rev. Lett. {\bf 86}, 5978 (2001).

\bibitem{yakiyama} M. Yakiyama, and Y. Hasegawa, Phys. Rev. B {\bf 67}, 014512 (2003).

\bibitem{scalapino} D. J. Scalapino,
Rev. Mod. Phys. {\bf 84}, 1383 (2012).

\bibitem{deguchi1} K. Deguchi, Z. Q. Mao, and Y. Maeno,
Phys. Rev. Lett. {\bf 92}, 047002 (2004).

\bibitem{deguchi2} K. Deguchi, Z. Q. Mao, H. Yaguchi, and Y. Maeno,
Phys. Soc. Jpn. {\bf 73}, 1313 (2004).

\bibitem{bradenpol} M. Braden, P. Steffens, Y. Sidis, J. Kulda, P. Bourges, S. Hayden, N. Kikugawa, and Y. Maeno
Phys. Rev. Lett. {\bf 92}, 097402 (2004).

\end{thebibliography}
\end{document}